%% file: main.tex
\documentclass[conference]{IEEEtran}
\IEEEoverridecommandlockouts
\usepackage{cite}
\usepackage{amsmath,amssymb,amsfonts}
\usepackage{algorithmic}
\usepackage{graphicx}
\usepackage{textcomp}
\usepackage{xcolor}
\def\BibTeX{{\rm B\kern-.05em{\sc i\kern-.025em b}\kern-.08em
    T\kern-.1667em\lower.7ex\hbox{E}\kern-.125emX}}
\begin{document}

\title{Educating for Hardware Specialization in the Chiplet Era: A Path for the HPC Community}

\author{\IEEEauthorblockN{Kazutomo Yoshii}
\IEEEauthorblockA{
\textit{Argonne National Laboratory}\\
Lemont, IL, USA}
\and
\IEEEauthorblockN{Mohamed El-Hadedy}
\IEEEauthorblockA{
\textit{California State Polytechnic University}\\
Pomona, CA, USA}
}

\maketitle

\input{abstract}


\input{narrative}

\section*{Acknowledgments}
This work is based on work supported by the U.S. Department of Energy, Office of Science, under contract  DE-AC02-06CH11357.


\bibliographystyle{IEEEtran}
\bibliography{ref}

\end{document}

%% file: abstract.tex
\begin{abstract}

The advent of chiplet technology introduces cutting-edge opportunities for constructing highly heterogeneous platforms with specialized accelerators. However, the HPC community currently lacks expertise in hardware development, a gap that must be bridged to leverage these advancements. Additionally, technologies like chiplet is cutting-edge with limited educational resource available. This paper addresses potential hardware specialization direction in HPC and how to cultivate these skills among students and staff, emphasizing the importance of understanding and developing custom hardware (e.g., rapid prototyping and resource estimation). We have been mentoring graduate-level students and new staff in hardware designs in a hands-on manner, encouraging them to utilize modern open-source hardware tools for their designs, which facilitates the sharing of research ideas. Additionally, we provide a summary of theses tools as part of our approach to prototyping and mentoring.


\end{abstract}

%% file: narrative.tex
\section{Introduction}

Hardware specialization has emerged as a promising approach for sustaining performance growth in the post-Moore era, given the increasing visibility of inefficiencies inherent in general-purpose architectures. 
Specialized accelerators (e.g., audio/video codecs, encryption engines), which can be found in system-on-chips, have proofread significant improvements in power, performance, and chip area over general-purpose architectures. Furthermore, emerging AI accelerators (e.g., Cerebras, Groq) are noteworthy examples of specialized architecture tailored to their target AI workloads.
In high-performance computing (HPC), most platforms widely employ heterogeneous node configurations incorporating CPUs and GPUs. However, as of today, no notable specialized accelerator integration has been found in HPC. While co-design activities between HPC laboratories and vendors exist, the primary objective is to design scalable platforms that support existing application portfolios.
Creating dedicated accelerators for specific workloads or applications has remained a secondary objective, which is understandable as adding new functionality to the chip can introduce more complexity to the entire chip development process, particularly verification.
Interestingly, industries are adopting chiplet technology~\cite{lau2021chiplet}, a modular design approach, to improve the yield, reusability, and scalability of their products. As chiplet technology matures and becomes more accessible technologically and economically, it could offer opportunities for the HPC community to evaluate custom, specialized accelerators tailored to our workloads. 

A 'what-if' question arises: if we could one day construct a new platform by combining multiple chiplets (many of which vendors offer common functions like memory blocks) without concerning many factors (e.g., license, technology accessibility, and economics), can we develop chiplet accelerators tailored to our specific workloads?~\cite{ yoshii2023hardware}
To this end, we must cultivate our expertise in hardware design, verification, and simulation, as well as our proficiency in system software stacks and compilers for custom accelerators. As it is challenging to hire hardware experts (e.g., shortage of hardware experts), we must be innovative in training existing staff and students whose major is computer science or domain science. While commercial EDA tools, IPs, and process design kits remain essential for chip designs with cutting-edge transistor nodes today, the emergence of open hardware standards like
RISC-V~\cite{asanovic2014instruction} and open-source hardware tools and designs is becoming increasingly pronounced.
For training and idea sharing purposes, we harness open-source hardware tools. These tools leverage modern software paradigms (e.g., functional programming, intermediate representation) and practices (e.g., test automation), making them particularly appealing to individuals whose expertise is in software, as well as accessibility (no license is required) extensively and possibly good community support.

\section{prototyping and mentoring}

At this point, we focus on mentoring students who actively participate in our projects rather than teaching a course. We provide design problems to students and work with them in a hands-on manner. Tools can often be showstoppers due to limited seat availability (commercial licenses), non-disclosure agreements, cost, and other barriers. To address these challenges, we have been evaluating open-source hardware tools for prototyping designs and mentoring, and we have successfully published several papers on hardware designs using these open-source tools~\cite{yoshii2022hardware, hammer2021strategies, strempfer2021designing, el2023reco, 10539908,
yoshii2023streaming, el2023bltesti}.

Here is a brief summary of the open-source hardware tools that we frequently use for prototyping and mentoring:

\textbf{Chisel Hardware Construction Language:}
Chisel~\cite{Bachrach:ft}, a domain-specific language for digital circuits, is implemented by harnessing the extensibility of the Scala programming language. 
Chisel has been used for many real-world tape-out chips~\cite{celio2017boomv2} (e.g., RISC-V processors, Google’s Edge TPU) and the Chisel community keeps growing. 
Chisel assumes positive-edge triggered registers and synchronous reset by default, resulting in more concise code, which works well for newcomers. 
Chisel has fewer dependencies, which is an important aspect of a teaching and prototyping tool. 
Various documents/tutorials and open-source codes written in Chisel are available online. 
Furthermore, Chisel is compatible with the Jupyter notebook~\cite{kluyver2016jupyter}, which aids in the teaching process.
We have also created a teaching material for internal purposes.


\textbf{Cocotb Verification Framework:}
Cocotb~\cite{rosser2018cocotb} is a verification library written in Python and converts user testbenches into an underlying simulator tool. It supports Icarus, Verilator, Synopsys VCS, and several other simulator tools. 
It has been gaining popularity among hobbyists, academia, and even industries~\cite{cocotb_survey_2023}.
Python ecosystems such as Numpy benefit test bench development and teaching (e.g., most of our input data are stored in a Numpy format).
Due to Python's popularity, introducing cocotb to software folks is relatively straightforward as long as the audiences have the basic concept of digital circuits (e.g., clock, wire, register). Cocotb can be installed using the pip tool, similar to other Python tools.

\textbf{OpenRoad:}
OpenRoad~\cite{kahngopenroad} is an electronic design automation (EDA) tool that transforms user's register-transfer level (RTL) codes, such as Verilog codes, into GDS, a database file format for data exchange of integrated circuit layout for fabrication.
We utilize OpenRoad primarily to obtain various reports from the EDA flow, such as die area, cell count, wire length, power, critical path, etc., to evaluate the physical aspect of the design instead of generating GDS for actual tape-out. 

\section{Conclusion}

We present a concise overview of three open-source tools that we have utilized for prototyping our own designs while mentoring students.
The list of open-source hardware tools can be extensive, especially for description languages like Chisel~\cite{truong2019golden}.
Other notable open-source hardware tools include CIRCT~\cite{eldridge2021mlir},
FireSIM~\cite{karandikar2018firesim},
and Chipyard~\cite{amid2020chipyard}.
Open-source hardware ecosystems are gaining momentum and hold potential as effective prototyping and training/education tools for Chiplet accelerator designs.
While we are still in the early stages of the educational aspect, we have gained expertise in mentoring using open-source tools and have developed design examples. We will continue working on creating a more concrete set of educational materials.